\documentclass{pasj01}
\usepackage{comment}
\draft 
\Received{$\langle$reception date$\rangle$}
\Accepted{$\langle$acception date$\rangle$}
\Published{$\langle$publication date$\rangle$}
\SetRunningHead{Kawachi et al.}{NIR brightening in a  Be-X-ray binary}

\begin{document}
\title{Near-infrared Brightening around the Periastron Passages of the Gamma-ray Binary PSR\,B1259$-$63 / LS\,2883}
\author{Akiko KAWACHI\altaffilmark{1}, Yuki MORITANI\altaffilmark{2}, Atsuo T.\, OKAZAKI\altaffilmark{3}, Hiromi YOSHIDA\altaffilmark{1} and Kenta SUZUKI\altaffilmark{1}}
\altaffiltext{1}{Department of Physics, School of Science, Tokai University, 4-1-1 Kita-kaname, Hiratsuka, Kanagawa 259-1292, Japan}
\altaffiltext{2}{Kavli Institute for the Physics and Mathematics
of the Universe (WPI), The University of Tokyo, 5-1-5 Kashiwa-no-ha,  
 Kashiwa, Chiba 277-8583, Japan}
\altaffiltext{3}{Faculty of Engineering, Hokkai-Gakuen University, 4-1-40, Asahi-machi, Toyohira-ku, Sapporo, Hokkaido 062-8605, Japan}
\email{kawachi@tsc.u-tokai.ac.jp}

\KeyWords{key word~${infrared:stars}_1$ --- key word~${stars:emission-line,Be}_2$ ---  key word~${binaries:close}_3$}

\maketitle

\begin{abstract}
The binary of the pulsar PSR\,B1259$-$63 and the Be star LS\,2883 has been observed at the 2010 and 2014 periastron passages in the near-infrared (NIR) bands using the IRSF/SIRIUS and SIRPOL. 
 The light curves in the {\textit J-},{\textit H-}, and {\textit Ks-}bands are almost identical 
 in these periastron passages.
 A flare starts  no later than 10 days before periastron and 
 the maximum brightening of about 0.1 magnitude is observed 
  12--17 days after periastron. 
 The rising part of the light curve is steeper and reaches a peak slightly 
 earlier in the {\textit Ks-}band 
  than in the other bands, thus a characteristic 
  track  appears 
  on the NIR color-magnitude diagram.  
  The  time lag between the NIR light curves 
 indicates that the variation in the Be circumstellar disk first occurs in an outer region.
  We propose that the initial rapid 
  contraction followed by the gradual expansion of the disk  is 
  evoked by the  rapidly changing 
  tidal torque around periastron and 
  the resultant change of the optically thick area causes the observed NIR light curves.
\end{abstract}

\section{Introduction} \label{sec:intro}
 PSR~B1259$-$63  is  a 48~ms period rotation-powered radio pulsar 
 in a highly eccentric ($e =\sim$0.87) and 
 wide ($P_{orb} =\sim$1236.72{\textit d) binary orbit \citep{2014MNRAS.437.3255S}
 with a Be star, LS~2883\citep{1992APJ.387.L37J}. 
  LS~2883 has parameters corresponding to an O9.5Ve star  
 whose mass $M_*$ is $\sim$31\,$M_\odot$ 
 and radius $R_*$ is $\sim$9.2$R_\odot$ \citep{2011APJ.732.L1N}.

 From the parallax measured by 
 the long baseline radio interferometric observations, 
 the distance to the binary is inferred 
 to be 2.6$^{+0.4}_{-0.3}$\,kpc\citep{2018MNRAS.479.4849M}.
 The parallax was also provided using an astrometric fit of 
  \textit{Gaia} Data Release 3  \citep{arXive:2012.03380} as 
  0.443 $\pm$ 0.013 mas which corresponds to an inversion distance 
  of 2.26 $\pm$ 0.07\,kpc.
  However, \citet{2018MNRAS.479.4849M} points out that 
 the semi-major axis of the orbit of LS\,2883 is subtended 
 of an order of 0.1 mas on the sky and the unmodelled orbital motion 
 could significantly affect the fitted astrometric parameters.
 These distance values are consistent with the result of optical spectral analysis of 
  LS\,2883 \citep{2011APJ.732.L1N} within the errors. 
 The pulsed emission becomes undetectable 
  around periastron ($\tau$) 
 during the period of about $\tau \pm$ 20{\textit d}
 \citep{2002MNRAS.336.1201C, 2014MNRAS.439.432C}).
 The absorption and the varying rotation and dispersion measures have shown that the Be circumstellar disk is inclined with respect to the orbital plane \citep{1995MNRAS.275.381M} and 
the pulsar is assumed to pass behind the 
 circumstellar disk around the periastron passage crossing the disk twice in 
 each orbit, before and after periastron.
 The binary separation is about 0.9 astronomical unit, 
 i.e. $\sim$20$R_*$ \citep{2011APJ.732.L1N} at its minimum  
 and is $\sim$45$R_*$  around the ending of pulse eclipse,  
  which implies that the Be-disk radius is larger than that 
 \citep{2016MNRAS.455.3674V}.

 The PSR\,B1259$-$63 system is 
 the first binary to be detected in the TeV gamma-rays \citep{2013AAPR.21.64D}.  The unpulsed X-ray emission is observed throughout the orbit 
\citep{1999APJ.521.718H}  and 
 the radio \citep{1999MNRAS.302.277J}, 
 X-ray \citep{1995APJ.453.424K,2006MNRAS.367.1201C}, 
 GeV \citep{2011APJ.736.L11A, 2015ApJ.811.68C}, 
 and TeV \citep{2005AAP.442.1A, 2009AAP.507.389A} 
  flares have been detected around periastron. 
 The multiwavelength observations have been repeatedly performed 
 (e.g. \cite{2014MNRAS.439.432C};~\cite{2020AAP.633.A102H}). 
 The two-peak flares in the radio and the X-ray 
 showed its first peak  at $\sim\tau-$15{\textit d} during the radio eclipse 
 and the second at $\sim\tau+$18{\textit d}, although the radio time profiles 
 varied
 orbit to orbit \citep{2005MNRAS.358.1069J}. 
 The X-ray spectrum is well described with a simple power-law with a photoelectric absorption and confirmed to extend up to 100~keV energy region 
\citep{2004AAP.426L.33S}.
  The broadband emission from the PSR\,B1259$-$63 system is 
 produced by high-energy particles in 
 the shocked pulsar wind.  
 For the radio, X-ray, and possibly, TeV gamma-ray bands,
 the orbital light curves around the periastron can be interpreted as a 
 result of the interaction of the pulsar wind with the stellar outflows of 
 the Be star (e.g. \cite{2014MNRAS.439.432C}).
The GeV flare shows puzzling features which start
 approximately 30{\textit d} 
 after periastron and have no counterpart in the other wavebands. 
 The mechanism of the GeV flaring episode is still under discussion 
 with  some theoretical interpretations considering either 
 the unshocked pulsar wind or Doppler-boosted emission from shocked material
 and/or the enhanced photon field other than LS\,2883 provides 
  at the quiescent orbital phase
  \citep{2012APJ.752L.17K, 2020MNRAS.497.648C, 2020AAP.633.A102H}.
Apart from these periastron events, an extended X-ray structure (clump)
 moving away from the binary was observed 
 by a series of \textit{Chandra X-ray Observatory} observations \citep{2015APJ.806.192P, 2019APJ.882.74H}.  The emission is considered as a high-speed ejecta 
 launched near periastron passage.  

 A Be star has a two-component extended atmosphere; 
 a polar wind of a low-density, fast outflow emitting UV radiation and 
an equatorial circumstellar disk 
  which is nearly Keplerian, consisting of a high density plasma from which the optical emission lines and the IR excess arise 
(e.g. \cite{2003PASP.115.1153P}).
Be disks in binary systems are considered to be subject to 
complex processes  such as precession, warping, tidal deformation and truncation \citep{2013AAPR.21.69R} and effects of the pulsar wind adds another complexity.
 \citet{2010PASJ.63.898O} and \citet{2012APJ.750.70T} simulated 
the PSR\,B1259$-$63 system 
using a framework of 3-dimensional Smoothed Particle Hydrodynamics (SPH) method. In their simulations, the pulsar wind disturbs the circumstellar environment around periastron.
The contact surface between the pulsar and stellar outflows 
 changes according to the density and geometrical parameters of the disk.  
 The disk under the influence of the tidal torque 
 also deforms density distribution and effective radial size according to the orbital phase.

 In order to understand the orbital perturbation on 
 the circumstellar disk of LS\,2883,  which can also be used as a probe to 
 constrain the high-energy emission mechanism of the binary, 
 we report the observation results of 
 the  LS\,2883 in the NIR band around the periastron passage.

\section{Observations and Analysis}\label{sec:obs}

The monitoring observations of the binary were performed around the 2010 and 2014 periastron passages (including the periastron $\tau$ of MJD\,55544.69 and 56781.42, respectively) 
for about a month in each orbital cycle 
with some observations in the quiescent phase ($\tau -$300{\textit d} -- $\tau-$85{\textit d} in  2010).

All the observations used the IRSF (InfraRed Survey Facility) 1.4 m telescope located at the Sutherland station of the South African Astronomical Observatory (SAAO) 
 with the 3-channel infrared camera, SIRIUS (Simultaneous InfraRed Imager for Unbiased Survey \citep{2003SPIE.4841.459N}) mounted on the Cassegrain focus of the telescope.
The SIRIUS camera is equipped with the two dichroic filters and the three 1024 $\times$ 1024 HgCdTe detectors (HAWAII arrays). 
It offers three  {\textit J}~(1.25~$\pm$~0.085~$\mu$m), {\textit H}~(1.63~$\pm$~0.15 $\mu$m), and {\textit Ks}~(2.14~$\pm$~0.16 $\mu$m) images simultaneously with a field of view of 7$\arcmin$.7 $\times$ 7$\arcmin$.7 and a pixel scale of 0\arcsec.45. 

We performed 156 observations in 25 nights in the 2010-cycle and 137 observations in  33 nights in the 2014-cycle. 
The typical seeing was 4-pixel for the {\textit H}-band, however, less photometric observations were included. 
A single image is typically combined with 25 dithered frames of 3 to 10-second exposures.  
For most of the 2010 observations (20 of 25 nights), the single beam polarimeter, SIRPOL(SIRius POLarimetry mode) was utilized. The SIRPOL consists of a half-wave plate rotator unit and a high efficiency wire-grid polarizer located upstream of the camera \citep{2006SPIE.6269.51K}. 
The typical accuracy of the polarization degree $\delta$P is $\sim$0.3\% depending on the stability of the sky. 

The primary data reduction has been carried out with the standard pipeline 
 software for the SIRIUS and SIRPOL\footnote{The pipelines are scripts based on IRAF command procedures:\\ 
https://sourceforge.net/projects/irsfsoftware/
} 
 including dark-subtraction, flat-fielding, self-sky subtraction, and combining of the dithered frames. 
The detectors of SIRIUS keep a good linearity ($<$~1\%) up to $\sim$10,000 ADU but saturate at $\sim$25,000 ADU, e.g. for the {\textit Ks}-band \citep{2014PASJ.65.27N}. 
Aperture photometry is followed using {\it IRAF APPHOT} task with the aperture sizes 
 adjusted with the FWHM's of multiple stars in the respective image. For the SIPOL data, the Stokes {\it I}-images are used for the photometric light curve. 
The SIRPOL {\it I} images and the SIRIUS images without the polarimeter taken in a short time interval are compared and the photometry results of these images are confirmed to agree well.

Differential photometry with the multiple reference stars are calculated and 
the 
light curves in reference to a quiescent phase observation ($\tau -$199{\textit d}) 
are deduced.  
The standard error of the zero point of instrumental magnitude is adopted in the error of each observation, together with the statistical errors.
 The deviations of the photometric results of a set of 20---30 sequential observations in a good-conditioned night are about 0.01 magnitude. 
The independent analysis of the observation at MJD$=$ 55549 ($\tau+$4{\textit d}) in the 2010 cycle   
 \citep{2012MNRAS.426.3135V} is consistent within the errors. 

\section{Results} \label{sec:results}
\subsection{Light Curves}

\begin{figure*}[!p]
\begin{center}
\begin{tabular}{cc}
\FigureFile(80mm,90mm){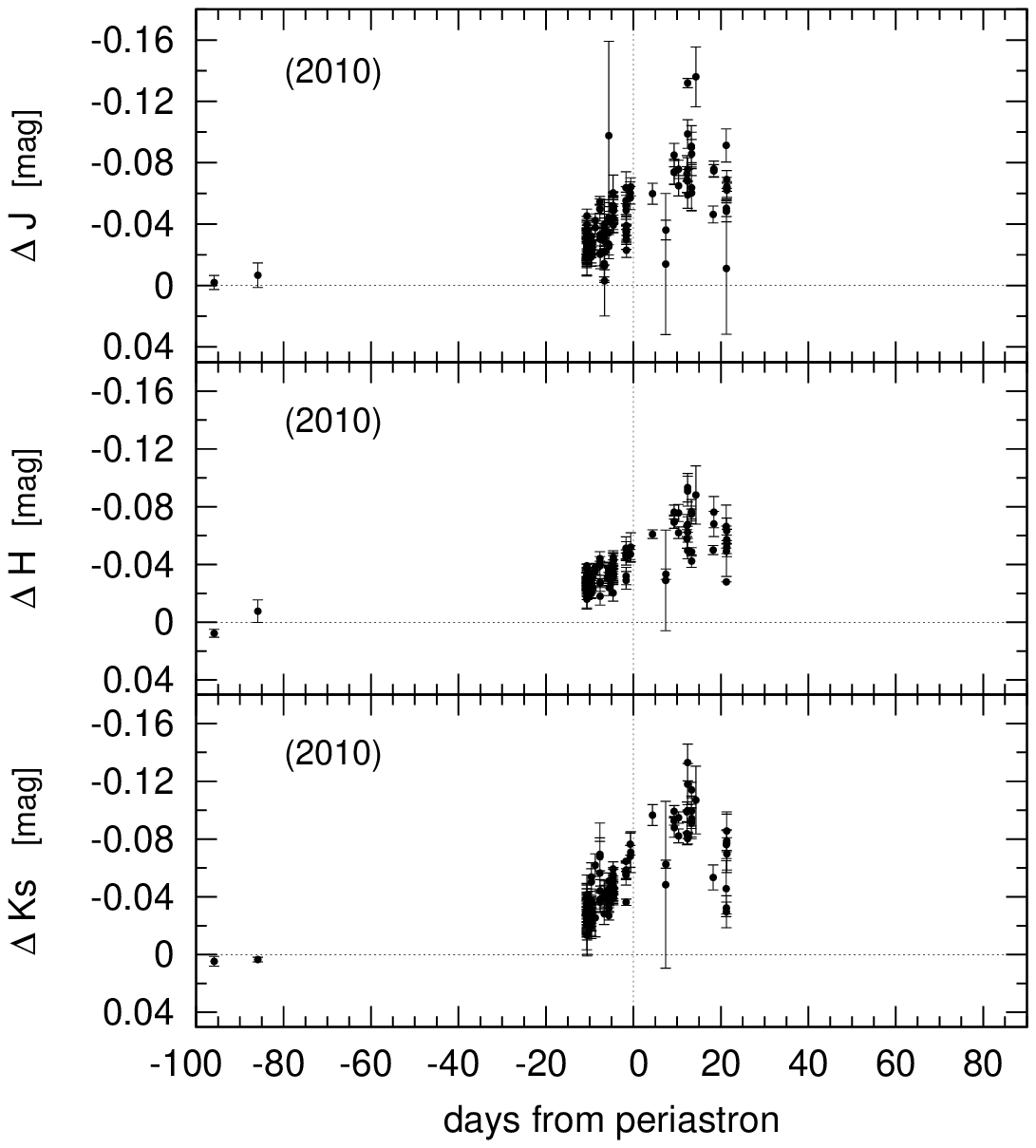} &
\FigureFile(80mm,90mm){./B1259NIRv2_fig1_right.eps} \\
\end{tabular}
\end{center}
\caption{The {\textit J}, {\textit H-}, and {\textit Ks-}bands 
  light curves as a function of days from periastron, for the 2010 cycle ({\textit left}) and the 2014 cycle ({\textit right}), respectively. 
 We use the observation in the quiescent phase ($\tau-$199{\textit d} in 2010) 
 as the common reference to the light curves in both orbital cycles. 
 (see the text).}
\label{fig:vsdays_side}
\end{figure*}

The differential light curves in the {\textit J-}, {\textit H-}, and {\textit Ks-} bands are shown in Figure\,\ref{fig:vsdays_side} as a function of days from the periastron.
In the 2014 orbital cycle, some data points of the {\textit J-} and {\textit Ks-} bands are missing due to the moderate data condition.  
 The flares of about 0.1 magnitude emerged in all the three bands 
  around periastron similarly in both of the orbital cycles.
 The flares reached the maximum brightness  about $\tau+$12{\textit d} and 
 gradually decayed afterwards. In 2014, the brightness was nearly back to the level in the quiescent phase 
  at $\tau+$70{\textit d}.
 The brightening episode might have started before 
 our first observation in the periastron passage ($\tau \sim -$10{\textit d}) but
 the significance is not high considering the errors. 
 In Figure\,\ref{fig:vsdays_overlap}, the  {\textit H}-band light curves for the 2010 and 2014 orbital cycles are overlapped.
 The flares of the two cycles 
 are almost identical except for  
 the slight differences in the timing of the maximum brightness and in 
 the decay time scale.

\begin{figure}[ht]
\centering
\FigureFile(80mm,100mm){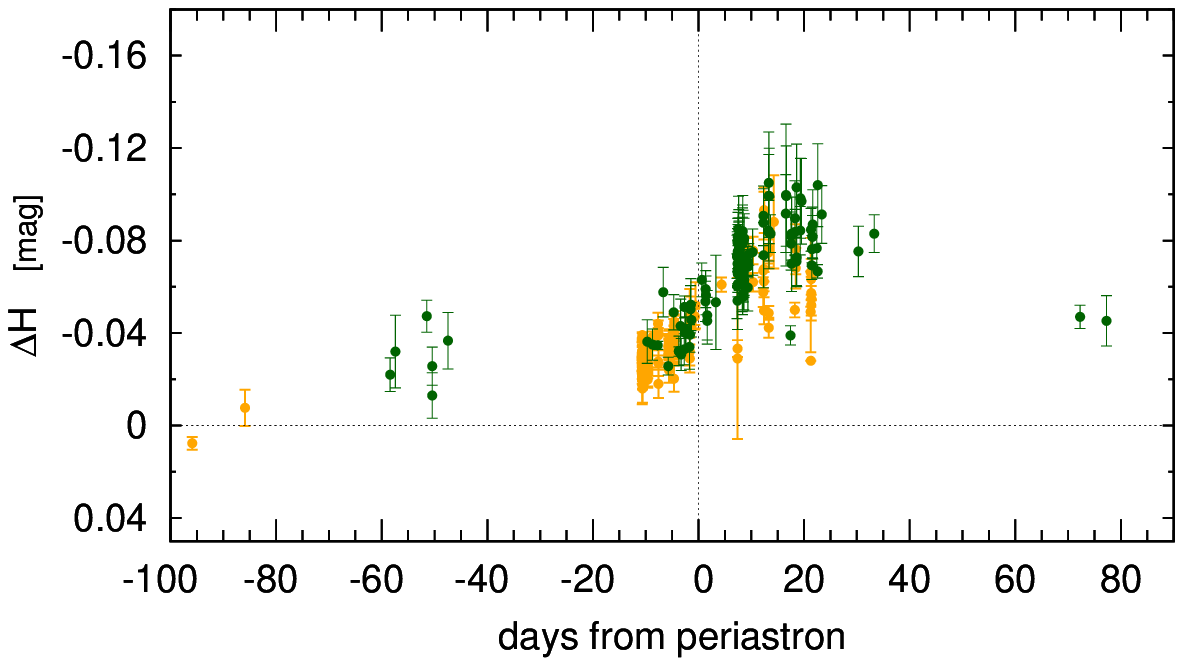}
\vspace*{.4cm}
\caption{The {\textit H}-band light curves in the two orbital cycles are 
 overplotted with the different colors; 
 the 2010 cycle (orange) and the 2014 cycle (green). }
\label{fig:vsdays_overlap}
\end{figure}

\subsection{Color-magnitude diagrams}

  The infrared color ({\textit H}$-${\textit Ks})
  and the {\textit H}-band magnitude  
 of the observation data are  plotted as diagrams for each orbital cycle 
 (Figure\,\ref{fig:color_mag_diagram}).
 The infrared color is calculated by taking 
 differential to the same reference observation in the analysis.
  The temporal changes of the flares in the {\textit J}-band and {\textit H}-band are about the same and 
 the  ({\textit J}$-${\textit H}) color values are approximately constant 
 over the observed phase.
 In these diagrams, different colors 
 correspond to different orbital phases. 
 The diagrams for the two orbital cycles exhibit a similar phase-dependent behavior.

  The infrared color goes redder  with  increase in the brightness until $\sim \tau+$10{\textit d} 
  then starts to turn blue 
  while the brightening still continues.
  At the flux maxima in the {\textit H}-band ($\sim \tau+$18d), 
 the color is almost back to that in the period prior to the flare. 
 This two-step color change is caused by the difference in 
  the temporal features of the brightening in 
  the {\textit Ks}-band and  the {\textit H}-band.
  The brightness in the {\textit Ks}-band increases more rapidly than in the 
  {\textit H}-band, which  makes  the infrared color redder at first, 
  ant the {\textit Ks}-band flux has a peak a few days earlier 
  than does the {\textit H}-band flux.
  However, the flux increase 
  of the {\textit H-}band gradually catches up with that of the 
 {\textit Ks}-band, which turns the color blue,  forming the upper branch 
 of the brightening track.
  The infrared color becomes redder again as the flare decays.
   The decay track in the diagram coincides 
  with the brightening track,  yet the kink phase 
  cannot clearly be located due to the lack of the  observations.

\begin{figure*}[!p]]
\begin{center}
\begin{tabular}{cc}
   \FigureFile(80mm,100mm){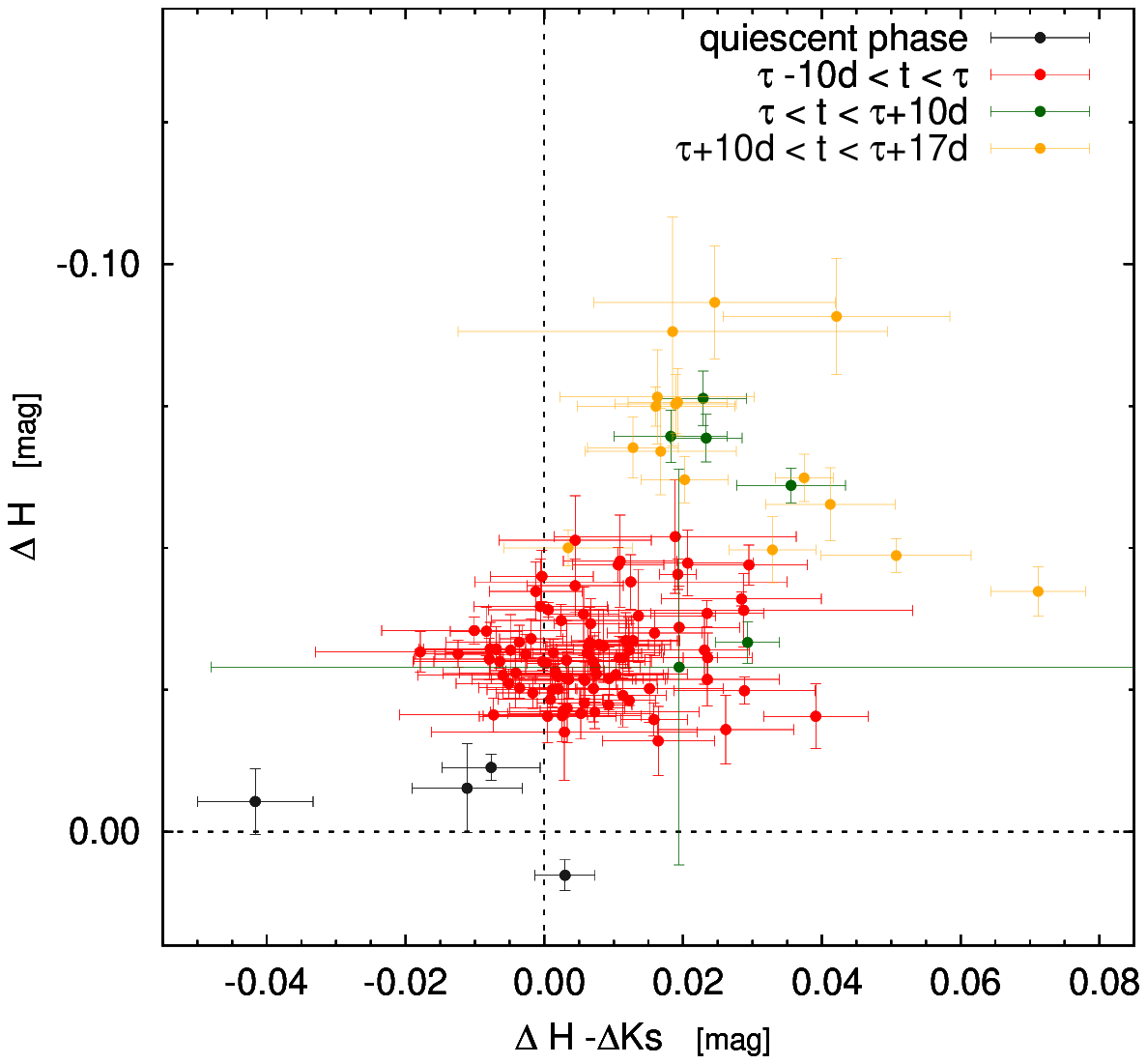} & 
   \FigureFile(80mm,100mm){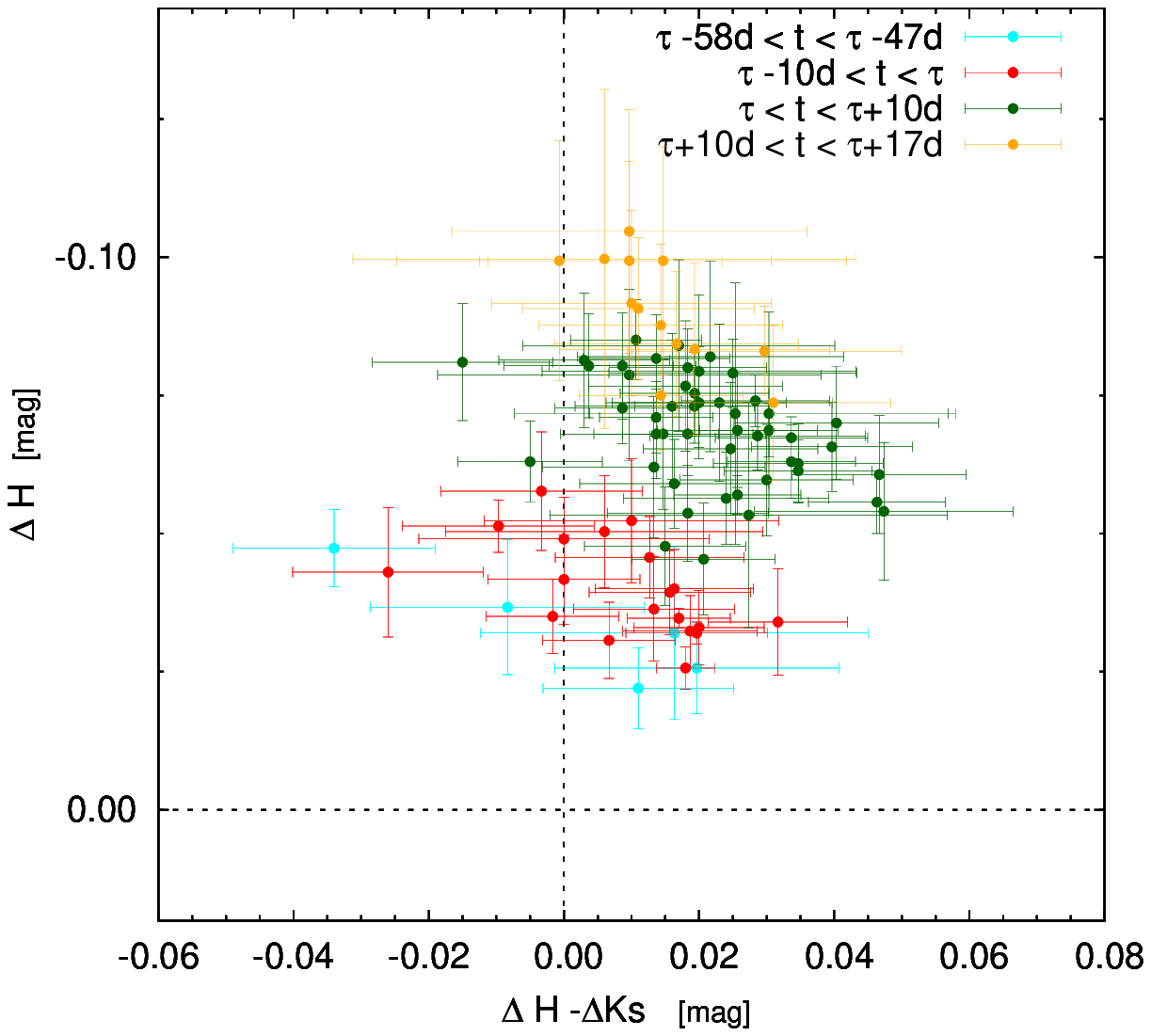} \\
   \FigureFile(80mm,100mm){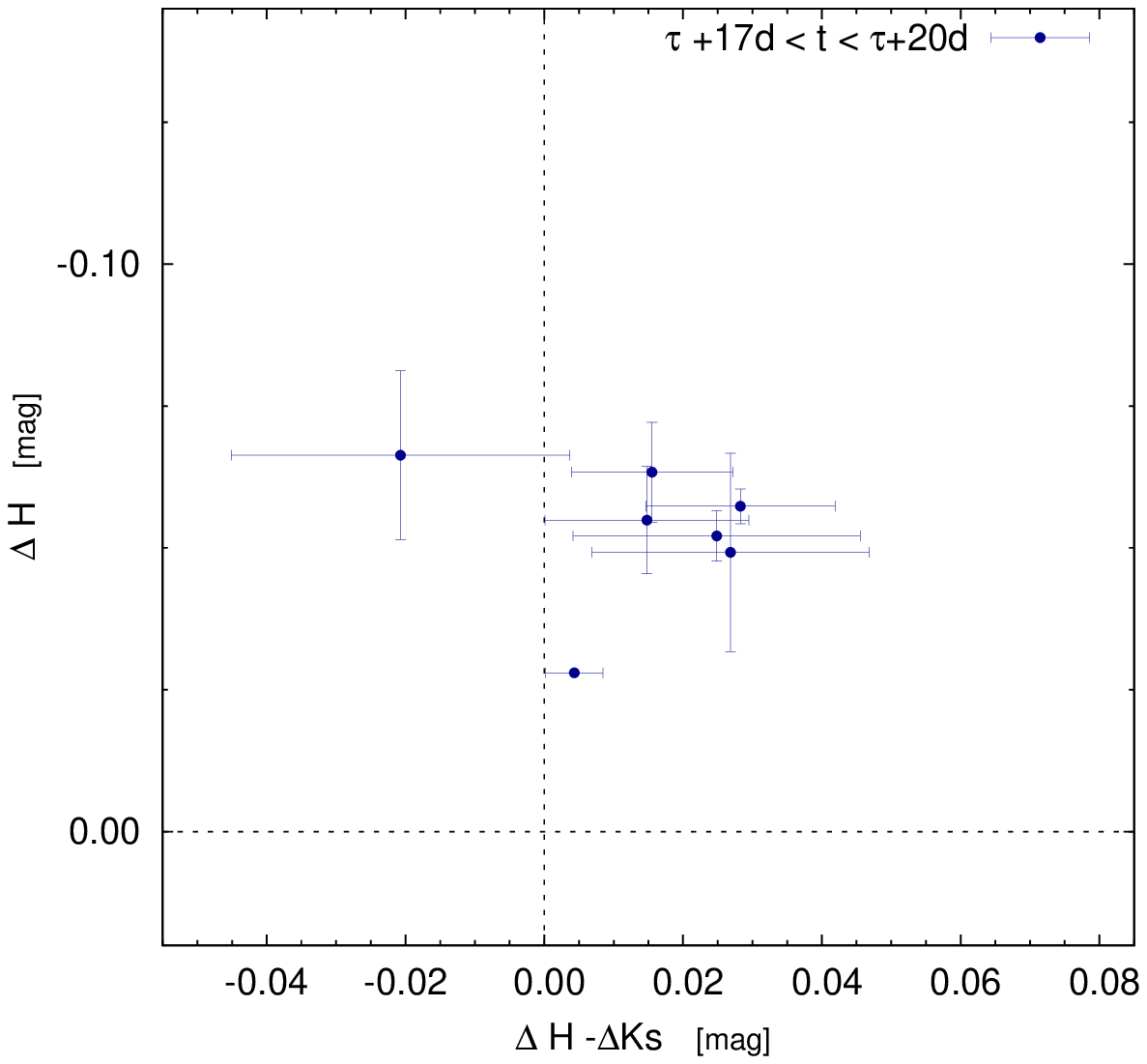} & 
   \FigureFile(80mm,100mm){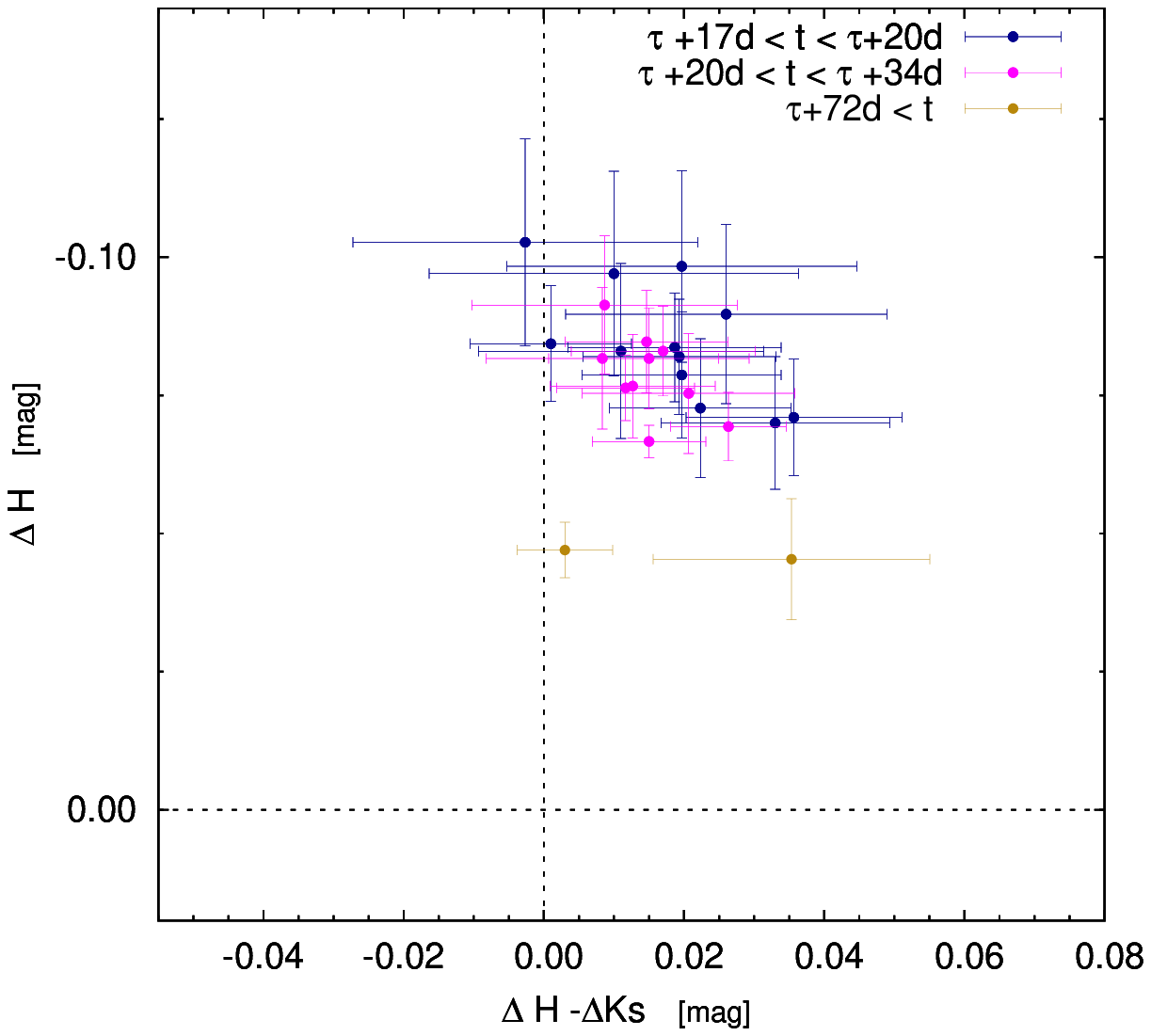} \\
\end{tabular}
\end{center}
\caption{
  The diagrams  of  the relative NIR color ($\Delta${\textit H} $-$ $\Delta${\textit Ks}) and the  { \textit H}-band differential magnitude $\Delta${\textit H} are 
 shown  for the periods of 
 brightening (top) and decay (bottom)  in the 2010 cycle ($left$) and the 2014 cycle ($right$), respectively. 
 Different colors of the points correspond to different orbital phases.
}
\label{fig:color_mag_diagram}
\end{figure*}

\section{Discussion} \label{sec:discussion}
\subsection{No association with the radio synchrotron flares}

  The brightening in the NIR band started no later than 10 days 
 prior to the periastron passage, 
  of which the timing is similar to that of the first flare in the unpulsed radio and X-ray emissions. 
 The flux of 1.4--8.4\,GHz increased from the detection limit 
 to several 10\,mJy. 
 Fitted by a power-law function of the form $S_\nu = C \nu^{\alpha}$, 
 the radio spectral index $\alpha$ is about $-$0.6 \citep{2005MNRAS.358.1069J}.
 If this simple power-law form is extrapolated from the radio to 
 the NIR band,  the contribution of this flare 
  would be  $\sim$10$^{-2}$\,mJy.
 This flux is about 4 orders of magnitude smaller than 
 the LS\,2883 NIR emission composed of the stellar atmosphere 
 and the circumstellar disk \citep{2011MNRAS.412.1721V}, by which the 
 total flux would vary only 10$^{-4}$ magnitude.
 We can, thus,  rule out the possibility of the association 
 of the radio synchrotron flare with 
 the observed NIR brightening of 0.1 magnitude of this work.

\subsection{Be-disk evolution by the tidal interactions and its effect on  the NIR emissions}
 The IR continuum excess of Be stars are interpreted as free-free 
 and free-bound emissions of the circumstellar disk \citep{2003PASP.115.1153P}.
 The IR emissivity of an optically thin disk 
 increases/decreases with increase/decrease in the local density.
  On the other hand, for optically thick disks,  the disk luminosity 
 is roughly proportional to the area where the disk is optically thick to the wavelength in consideration.
 The radius of the optically thick region is larger for a longer wavelength. 
  For more details see Figs.\,2 and 7 of \citet{2013AAPR.21.69R}.

  The tidal effect of the neutron star on the Be disk is enhanced
  around periastron in eccentric Be binaries. \citet{2002MNRAS.337.967O}
  has studied the orbital modulation of the Be disk structure as well as
  the long-term disk evolution  under the influence of the neutron star
  using a three-dimensional SPH code.
  In the snapshots at several orbital phases (Figs 10 and 11 of \citet{2002MNRAS.337.967O}),  
 the disk starts contraction just before periastron by the tidal torque from the neutron star 
 and then expands as the neutron star moves away from periastron.
  Thus the effective disk radius modulates around the mean value.
  Throughout this period of strong tidal interaction, the disk is deformed. After repeating 
   contraction and expansion a few more times, 
  the disk becomes almost axisymmetric and 
  starts to restores its radius slowly by viscous diffusion.
  Although the strength of the interaction 
  is dependent on the  inclination angle  of Be-disk to the orbital plane, 
  this variation of the disk structure, in density distribution and 
 in effective radius, is  expected to be observed 
 around the binary periastron passage.

Figure\,\ref{fig:combined_w_He_I} shows the optical spectroscopic monitoring results of the PSR\,B1259$-$63 binary at the 2014 periastron passage 
  \citep{2016MNRAS.455.3674V}  
 together with the {$H-$}band differential light curve and NIR color variation of this work.
 The He$_I~(\lambda$~6678) line shows a consistent double peak structure 
 but showed variation in the strength.
 The constraint on the emission location of He$_I$  (Fig.\,\ref{fig:combined_w_He_I} (b))  is  derived as in \citet{2016MNRAS.455.3674V}   
from the peak separation of the lines 
  assuming a Keplerian disk. They use 
 $v~{\rm sin}~i = 260 \pm 15 {\rm km s}^{-1}$  \citep{2011APJ.732.L1N} in the calculation.  
 The He$_I$ location moves outwards after periastron 
  and reaches a maximum radius around $\tau+$14{\textit d} 
  when the NIR bands were 
  brightest (Fig.\ref{fig:combined_w_He_I} c). 

 The movement of the location of the He$_I$ emission can be interpreted by the 
  expansion of the disk after 
 the contraction in the inner part evoked by the tidal torque.
  Assuming the LS\,2883 disk is optically thick, 
  the expansion of the disk leads to an increase of the optically thick area 
  and hence of the NIR flux. Therefore it is no coincidence 
 that the NIR flares  are observed when 
  the He$_I$ emission location goes outwards.
 The increase of the EW of H$_\alpha$ line \citep{2015MNRAS.454.1358C, 
 2020MNRAS.497.648C} is qualitatively 
 consistent with this interpretation. 
 Here, we note that it is rather natural that 
 the initial tidal contraction is not seen in the light curve of the H$_\alpha$ line, given that the line mainly arises from
 the outer part of the disk, where the pulsar wind has a much 
 stronger effect than the tidal torque.
  The He$_I$ location might have shifted inwards 
  around $\tau-20${\textit d} as a sign of disk contraction, although 
  the shift is not significant given the size of the error bars.
  There is also a hint of oscillation around $\tau+40${\textit d}, whic  is neither statistically significant.
  The V/R, the ratio of the height of the V component to 
  that of the R component, of the double-peaked He$_I$ line has local maxima  around 
  $\tau-20${\textit d} and $\tau+40${\textit d}. 
This variation suggests the 
 strong disk deformation occurs around these dates, 
which are close to the  timings of 
 the first disk crossing of the pulsar and 
 the GeV flare, respectively.
Unfortunately, the NIR observations are missing in these periods.

  As described in Section\,\ref{sec:results}, 
  the observed kinked pattern on the NIR color-magnitude diagram is  
   the result of the different timings of 
  brightening between the {\textit H-} and {\textit Ks-}bands.
  The earlier response in the {\textit Ks-}band indicates that 
  the disk variation starts in the outer region.  
  In the  PSR\,B1259$-$63 / LS\,2883 binary, 
  the Be disk expands 
  first in the outer region and then in the inner region.

  The destruction of the LS\,2883 Be disk has been discussed 
  to explain the trigger of the GeV flaring 
\citep{2015MNRAS.454.1358C} 
 or the extended X-ray ejecta \citep{2015APJ.806.192P}.
 Such an event is most likely to occur in the outermost region of the disk, where the 
 pulsar wind effect is strongest. On the other hand, the NIR variability studied in this 
 work arises from an inner intact region and cannot be directly connected with 
 the disk destruction episodes.

  \citet{2016MNRAS.461.2616P} calculated light curves for 
  $\lambda =$ 0.5 $\mu$m, 3$\mu$m, and 1\,mm  using  the SPH simulation data of a Be-star binary. 
  Their calculation shows the brightening occur after periastron at 
  different timings for different wavelengths. However, 
  because of the simulation setting where 
  the compact object moves within the disk from well before  periastron, 
   a direct comparison with our results is difficult.

\begin{figure}[ht]
\centering
\FigureFile(80mm,100mm){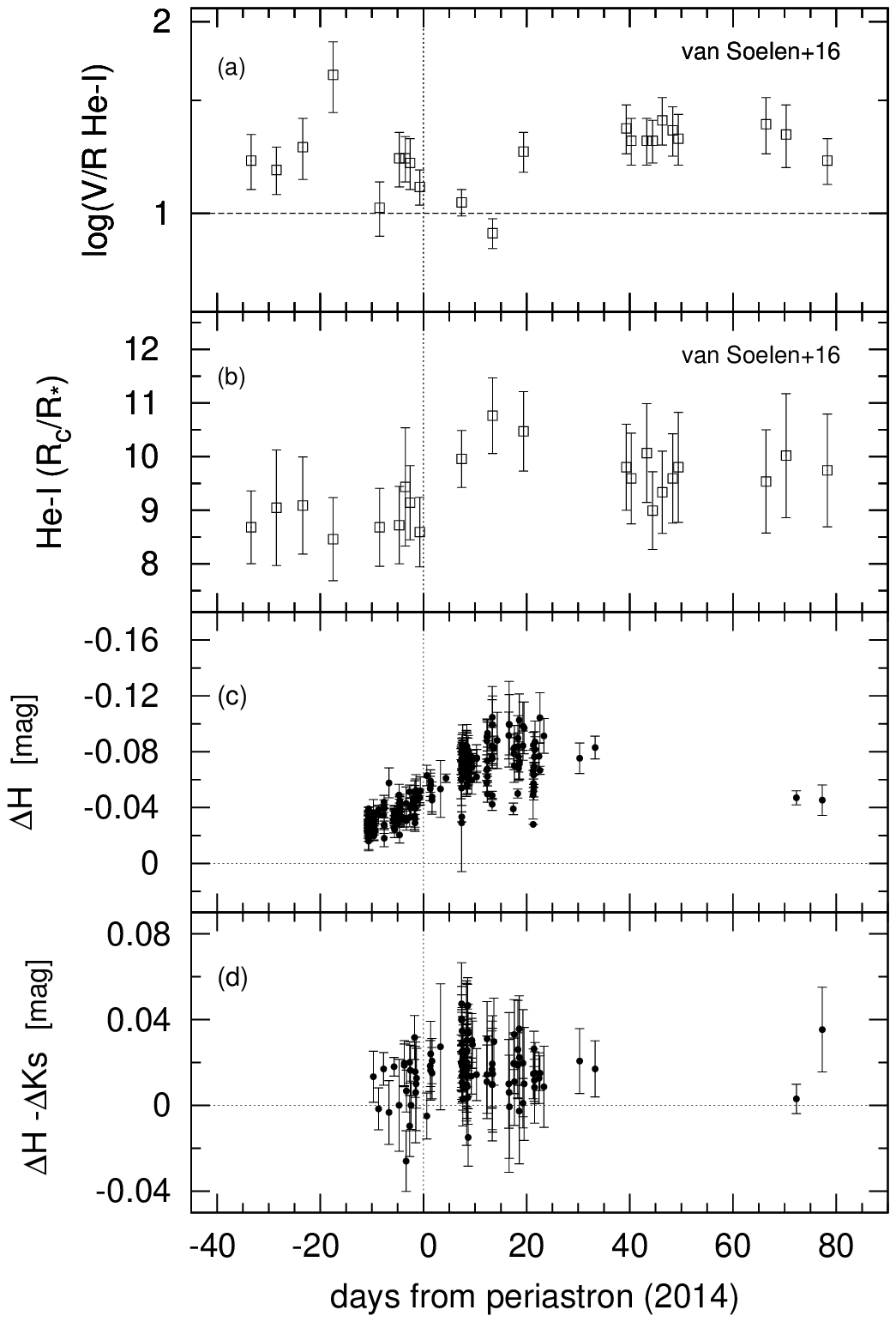}
\caption{Variations in the  He$_{\rm I}$ emission \citep{2016MNRAS.455.3674V}
 in the 2014 orbital cycle of the binary 
 are compared with the NIR observations of this work. 
 (a) VR ratio of the He$_{\rm I}$ line emission  shown in log scale,  
 (b) location of the He$_{\rm I}$  emission R$_C$  in a unit of stellar radius R$_*$, 
 (c) differential light curve in the {\textit H}-band, 
  $\Delta${\textit H}  and 
  (d) relative NIR color variation ($\Delta${\textit H} $-$ $\Delta${\textit Ks}).
}
\label{fig:combined_w_He_I}
\end{figure}

\section{Conclusions}
 The NIR monitoring of PSR\,B1259$-$63 was performed around 
 the 2010 and 2014 periastron passages. 
    Almost identical flaring light curves with the maximum brightening 
 of about 0.1 magnitude 
 in the {\textit J}, {\textit H-}, and {\textit Ks-}bands were observed 
 in  both cycles.  
  Associated with the flaring event, 
 a characteristic kinked pattern appeared on the 
   NIR color-magnitude diagram.
  This phase-dependent movement on 
  the diagram is due to the fact 
  the {\textit Ks}-band brightens faster than the other two bands, 
  which indicates that 
  the variation in the Be circumstellar disk is first 
  excited  in the outer region and then propagates inwards.
  The observed NIR variability is likely caused by 
  the Be disk expansion and the resultant increase of optically thick areas, 
  which is evoked by the tidal interaction  with the pulsar 
 around  periastron.

\chapter*{Acknowledgments}
This work is supported by JSPS KAKENHI 
 Grant Numbers JP21540304, JP20540236, and JP24540235, 
 and 
 National Research Foundation 
 under the Japan-South African Research Cooperative Program.

\end{document}